# Solving the $CH_4^-$ riddle: the fundamental role of spin to explain metastable anionic methane

Alejandro Ramírez-Solís[1*], Jacques Viguè[2], Guillermo Hinojosa[3], and Humberto Saint-Martin[3]

[1]Centro de Investigación en Ciencias-IICBA, Universidad Autónoma del Estado de Morelos. Cuernavaca, Morelos 62209. México.

[2]Laboratoire Collisions Agrégats et Réactivité-IRSAMC, Université Paul Sabatier and CNRS UMR 5589. 118, Route de Narbonne. 31062 Toulouse Cedex, France.

[3]Instituto de Ciencias Físicas, Universidad Nacional Autómoma de México. Cuernavaca, Morelos 62210 México.

Abstract

When atoms or molecules exist in the form of stable negative ions, they play a crucial role in the gas phase chemistry. Determining the existence of such an ion, its internal energy and its stability are necessary prerequisites to analyze the role of this ion in a particular medium. Experimental evidence of the existence of a negative methane ion $CH_4^-$ has been provided over a period of 50 years. However, quantum chemistry had not been able to explain its existence, and a detailed recent study has shown that the experimentally observed species cannot be described by the attachement of an electron in the ground state of $CH_4^-$. Here we describe $CH_4^-$ as being a metastable species in its lowest quartet spin state and we find that this species is a $CH_2^-:H_2$ exciplex with three open shells, lying 5.8 eV above the methane singlet ground state but slightly below the dissociation fragments. The formation of charged exciplexes is a novel mechanism to explain small molecular anions with implications in a plethora of basic and applied research fields.

e-mail:alex@uaem.mx

## The $CH_4^-$ puzzle

Methane has 10 electrons in a noble gas-like configuration so that electron capture by this very stable molecule is considered impossible. Nonetheless, electron attachment in methane has consistently been evidenced by several groups over five decades [1-10] though the true nature of this anionic species still remains unknown. Therefore, the elucidation of the origin and nature of anionic methane would be an important finding with implications on several fields, like our current understanding of atmospheric chemistry, hydrocarbon plasma, flame science, cluster science, space physics, planetary science and quantum chemistry.

Molecular hydrocarbon anions are called upon to explain charge distributions in plasma and to interpret spectroscopic observations. Based on available theoretical models and emission spectroscopic experiments, molecular anions have been detected in plasma atmospheric environments such as in a comet's coma, Titan's ionosphere, or in the interstellar medium. Molecular anions are believed to have important roles in the formation of aerosol dust in the chemical composition of Titan's atmosphere and in plasma plumes. These cold plasma are extremely complex and the explanation of their behaviour frequently resorts to negative molecular species [11]. For instance, in a simulation of Titan's atmosphere, Horvart *et al.* [12] propose negative molecular anions resulting from rather complex dissociative electron attachment channels to interpret the charge distribution in a $N_2 - CH_4$ plasma. Despite the fact that methane is the most abundant species, they did not consider negative methane to be present in their plasma; perhaps this is because $CH_4$ is believed to be incapable of capturing an extra electron. However, $CH_4^-$ would provide a much simpler explanation for the presence of metastable molecular anions.

Another field where the confirmation of this anion would be an important discovery is quantum chemistry. Dipole-bound anion formation is the most accepted mechanism to explain the capture of an extra electron by neutral molecules. This mechanism explains the formation of several molecular anions that are important in astrophysics [13,14]. Since methane has zero dipole moment, the confirmation of $CH_4^-$ challenges the dipole-bound model as the only mechanism for the formation of simple molecular anions. We present here a new quantum mechanical mechanism capable of explaining the observed methane anionic species.

The first observations of electron attachment in methane were reported by Trepka and Neurt [1] and by Sharp and Dowell [2], who used a beam technique and found that dissociative electron attachment in $CH_4$ took place over the 8 - 13 eV range of electron energies, with $H^-$ and $CH_2^-$ as dissociative product ions. Later, using a pulsed Townsend technique, Hunter *et al*. [4] measured the density-normalized electron-attachment coefficient for methane over the geometry independent field strength, E/N, in the 52.5 - 250 Td range. Electron-impact ionization processes have been thoroughly studied for gaseous methane over a very wide range of the density-normalized electric field strength up to $5 \times 10^4$ Td, also suggesting the presence of the methane anion [6]. In Townsend measurements at low E/N [8] large amounts of heavy negative ions were clearly observed so that the electron-attachment process in $CH_4$ is certain, thus providing irrefutable evidence of the presence of the methane anion [10].

From the theoretical side, the electron affinity (EA) was calculated as it is usually done [15], i.e., as the difference between the energy of the doublet ground state $^2A_1$ of the negative ion minus the energy of the singlet ground state $^1A_1$ of neutral methane plus a free electron. The extra electron in the ground doublet state of $CH_4^-$ goes into a quasi-spherically symmetric Rydberg molecular orbital (described as mainly carbon 3s) [16]. It was found that the EA of methane was large, ranging from 6.1 to 1.2 eV using second and fourth order Möller-Plesset perturbation theory (MP2 and MP4) and even applying variational Configuration Interaction methods (QCISD) with a variety of gaussian basis sets ranging from 6-31G(d,p) up to 6-3111++G(3df,2p). When the complete basis set (CBS) extrapolation [17] was applied, the EA was estimated to be 0.5 eV. However, while the decrease of the EA with basis set quality is clear, none of the electronic structure methods used the more accurate correlation-consistent polarized valence (cc-pVnZ) basis sets [18].

From the electronic structure point of view, the fact that the negatively charged $CH_4^-(^2A_1)$ species will eventually produce $CH_4(^1A_1)$ plus a free electron means that, for a given level of theory, the energy of the ground doublet state of the anion should approach that of methane as the quality of the atomic basis sets improves. This stems from the fact that larger and more diffuse basis sets allow the extra quasi-spherical electron to move away from the neutral tetrahedral molecule, which has no dipole moment capable of inducing an attractive charge-dipole interaction, thus lowering the total energy of the ($CH_4$, e-) pair at infinite distance. Therefore, a recent study addressed the evolution of the electron affinity of methane with highly correlated benchmark *ab initio* calculations with Dunning's aug-cc-pVnZ basis sets up to aug-cc-pV6Z + diffuse [18]. The electron affinity was calculated at the MP2 and Coupled Cluster [CCSD(T)] levels including extrapolations to the CBS limit [19]. This study showed that the energy of the $^2A_1$ state of the anion asymptotically approaches that of the $^1A_1$ ground state of $CH_4$ plus a free electron with increasing basis set quality. The inescapable conclusion of that work is that an excited electronic state of anionic methane must be at the root of the experimental observations of $CH_4^-$.

The comparison of the experimental evidence and these quantum-chemical results suggests that the detected negative ion is not correlated to a ground state ion but to a metastable negative ion. This type of negative ion is well known for atoms, $He^-$ being the most striking case, but many other examples are known, as reviewed in the paper by Bunge *et al*. [20]. The case of $He^-$ has been much studied since its discovery [21]: $He^-$ is in the 1s2s2p ($^4P_J$) levels, with an electron binding energy equal to 77 meV with respect to the 1s2s ($^3S_1$) first excited state. The lifetimes of the $^4P_J$ levels have been calculated and measured, with the J=5/2 level having the longest lifetime, with an experimental value of 343±10 µs, in agreement with the theoretical prediction [22]. Methane is isoelectronic with neon and, following Bunge *et al*. [20], neon has no metastable negative ion. However, $CH_4^-$ is isoelectronic with sodium and Feldman and Novick [23] showed that the alkalis have long-lived quartet states which are in the ionization continuum and metastable with respect to ionization. These quartet states have large internal energies, varying from 57.3 ± 0.3 eV for the lithium 1s2s2p ($^4$P) level to 12.6±0.3 eV for the potassium $3p^54s3d$ ($^4$F) level (zero energy corresponds to the atomic ground state). The lifetime of these states are surprisingly long for auto-ionizing states: if auto-ionization was fully allowed, their lifetimes should be of the order of $10^{-15}$s to $10^{-13}$s, while the observed lifetimes vary from 5.1±1 μs for lithium to 90± 20 μs for potassium [24]. There is no measurement of the lifetime of the sodium quartet state $2p^53s3p$ ($^4$D) but a theoretical evaluation of the decay rates of the quartet states of sodium has been made by Holmgren *et al*. [25]. It is interesting to note that three levels, namely the $2p^53s3p$ ($^4D_{7/2}$), the $2p^53s4s$ ($^4P_{5/2}$) and the $2p^53s3d$ ($^4F_{9/2}$) levels, have negligibly small auto-ionization rates and the decay of these levels is dominated by radiation emission toward states of the quartet symmetry, the lowest quartet states having very long lifetimes.

The exceptional properties of these high-lying metastable states of the alkali atoms has given us the idea that a methane negative ion could also be described by the spin quartet symmetry. As in the study by Holmgren *et al*. [25], one may expect that some quartet states have sufficiently small auto-ionization and radiative decay rates to explain the observed $CH_4^-$ species. Thus, two key questions arise. First, what is the energy of the lowest quartet state of the $CH_4^-$ ion with respect to the ground state of $CH_4$? and, secondly, is the resulting anionic species stable enough for it to be detected in experiments at room tempereature? We address these issues in two ways. Firstly, through benchmark *ab initio* quantum chemical calculations, including both non-dynamic and dynamic electronic correlation effects with the Complete Active Space Self-Consistent Field (CASSCF) and the second order MP2 scheme with the CBS extrapolation. Secondly, but crucially aimed at explaining experimental evidence, we focus on the dynamical behaviour of the excited quartet methane anion at room temperature through Born-Oppenheimer molecular dynamics (BOMD) based on Density Functional theory (DFT) using a calibrated hybrid exchange-correlation functional. Computational details can be found in the Supplementary Material.

**The structure of $CH_4^-$ at 0 K**

We address the energy and the structure of the $CH_4^-$ species in its lowest quartet (S=3/2) spin state by fully unrestricted MP2/aug-cc-pVnZ optimizations to obtain the CBS energy limit with the n = 3,4,5 series; the lowest quartet state energies are relative to those of the neutral methane molecule at a given level of theory. We explored the possibility that the lowest quartet state might have some multireference character through CASSCF(7,7)/aug-cc-pVnZ calculations; we found that this is not the case so that the single-reference MP2 approach yields reliable results. Geometry optimizations at the MP2 and CASSCF levels lead to the same molecular picture where a weakly bound $CH_2^-$:$H_2$ exciplex is produced upon relaxation of the excited anionic methane. The energies of the separate optimized fragments were also determined at the MP2/aug-cc-pVnZ level, leading to our best calculated CBS value (1.2 kcal/mol) for the dissociation of the exciplex into $CH_2^-$(S=3/2) + $H_2$(S=0).

Since the thermal energy at 300 K is 0.59 kcal/mol, at the highest level of theory (MP2/CBS limit), this well will hold the exciplex together at room temperature, and this remains true even neglecting equipartition of the thermal energy amongst all the vibrational modes of the exciplex. Energies, the optimized geometry, charge and spin distributions can be found in the Supplementary Material.

**Stability of the $CH_2^-$ :$H_2$ exciplex at 300 K**

Since the experimental detection of negative methane has been done at room temperature, we analyze the behaviour of the lowest excited quartet state at 300 K through Born-Oppenheimer molecular dynamics. We utilized the same BOMD approach [26] as that reported in [27].

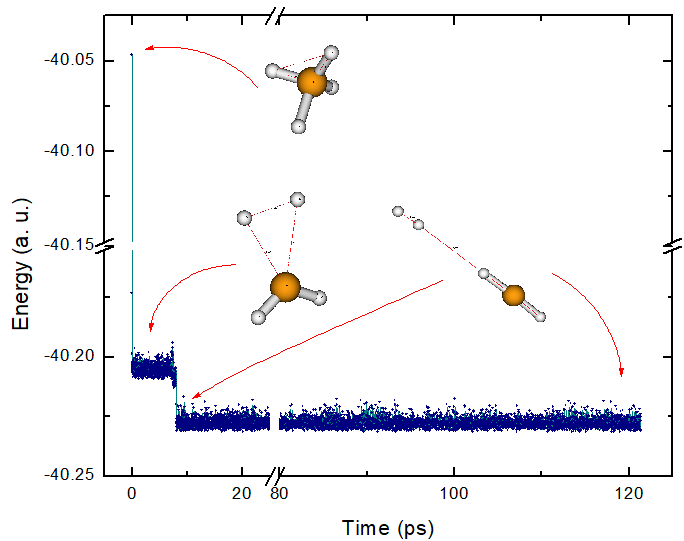

Figure 1: Energy evolution of the excited quartet state of $CH_4^-$ at 300 K. The deep about 8 ps indicates the appearance of the $CH_2^-$:$H_2$ exciplex. Metastability with respect to the separate $CH_2^-$ (S=3/2) + $H_2$ (S=0) fragments was found over an extended time of, at least, 1 ns at 300 K.

Before starting the BOMD simulations we tested several GGA, meta-GGA and hybrid exchange-correlation (XC) density functionals to find which one is capable of reproducing the CASSCF and MP2 optimization results, i.e., we verified that it does indeed lead to the $CH_2^-$:$H_2$ exciplex at 0 K. This calibration revealed that the hybrid B3LYP functional qualitatively reproduces the CASSCF and MP2 picture for the lowest quartet state. Then, 1 ns BOMD B3LYP-based simulations ($2 \times 10^6$ configurations) were performed on the $CH_4^-$ system imposing a total spin of 3/2 (details in Supplementary Material). The optimized $T_d$ geometry of neutral methane was used as starting point with no preferred nuclear velocities other than those obtained by a Boltzmann distribution at 300 K. Figure 1 shows the energetic and structural evolution of anionic methane in its lowest spin quartet state. Note that a nearly linear stable $CH_2^-$:$H_2$ excited complex is formed in the first 10 ps.

The infrared spectrum was obtained from the Fourier transform of the velocity autocorrelation function after thermalization was achieved (10 ps) at 300 K and is shown in Figure 2.

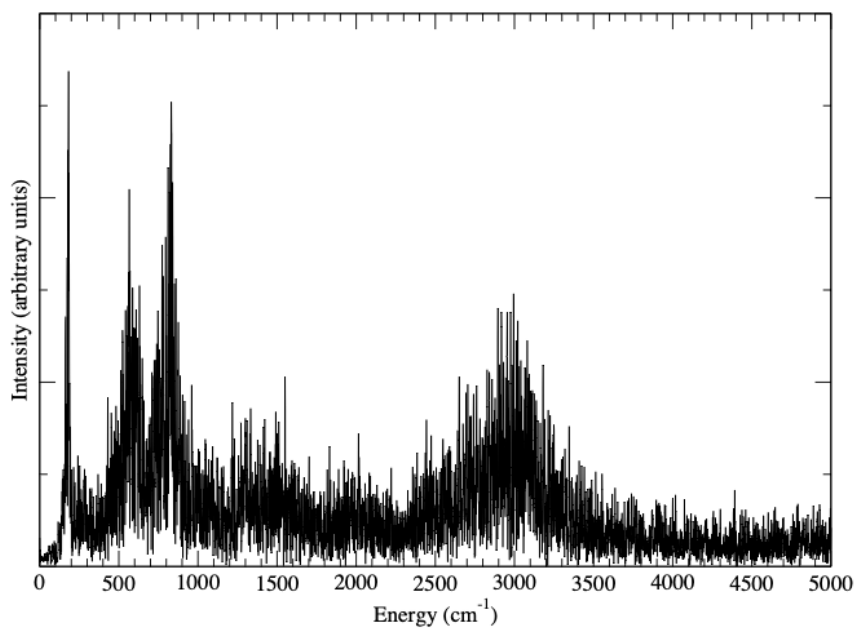

Figure 2: Vibrational spectrum of the excited quartet state of $CH_4^-$ leading to the $CH_2^-{:}H_2$ exciplex.

The $CH_4^-$ spectrum should be compared to the vibrational spectra of the separate $CH_2^-$ + $H_2$ species at the same level of theory, as shown in Fig. SM-4XX. The isolated $CH_2^-$(S=3/2) anion has three modes, two stretching at 923 and 1600 cm$^{-1}$, and one bending mode at 399 cm$^{-1}$; the vibrational mode of $H_2$ is at 4410 cm$^{-1}$. The $CH_4^-$ spectrum clearly reveals the weakly bonded nature of this anionic species since the inter-moiety stretching mode is found at 180 cm$^{-1}$. The lowest frequency peak is associated with the weak bond of the $CH_2^-$ - $H_2$ units, but the present results at 300 K shows that the lifetime of this anionic species is longer than 1 ns and many experiments suggest that metastability might exist for time scales three to four orders of magnitude larger at room temperature.

### Conclusions and perspectives

Electron attachment in methane has been detected by several groups over several decades. However, the ground doublet state of anionic methane has been shown to be unstable with respect to $CH_4$ plus a free electron through benchmark *ab initio* calculations [19]. Thus, some excited state must be at the root of the observed $CH_4^-$ species which has a longer lifetime than the time needed (ca. $10^{-6}$ s) to reach the detection apparatus. Methane is isoelectronic with sodium and, based on the exceptional properties of the spin = 3/2 high-lying metastable states of the alkali atoms, we have explored the idea that a methane negative ion could also be described by the spin quartet symmetry. We performed benchmark *ab initio* CASSCF and MP2 calculations of the anionic system in the lowest quartet state. Both types of methods lead to a weakly bound $CH_2^-{:}H_2$ exciplex. This high-spin exciplex explains, at the same time, the observation of anionic methane but also the detection of the $CH_2^-$ species. The thermal stability of this exciplex was studied at 300 K through Born-Oppenheimer molecular dynamics simulations based on a calibrated exchange-correlation functional. Although these quantum chemical simulations address a time scale below the upper bound of µs in the experiments, we found thermal stability of the $CH_2^-$:$H_2$ exciplex at 300 K with a total spin of 3/2 for at least 1 ns, providing an explanation for some of the observed anions. The infrared spectrum predicts a peak *ca*. 180 cm$^{-1}$ that, if detected, can be interpreted as the signature of the $CH_2^-$-$H_2$ stretching mode of the exciplex.

We propose a bold and challenging molecular picture that explains the presence of anionic methane in many experiments. The present study shows that when electronic attachment occurs in $CH_4$ molecules, the spin quantum number of some of the resulting anions is 3/2 instead of 1/2, as one could naively expect, revealing the pivotal role played by the lowest quartet state of this elusive anionic molecular species. These results call for new Stern-Gerlach experiments or the infrared spectrum of the m/q = −16 molecular beam in future investigations. Further experimental data on the lifetime or refined electron affinity measurements of anionic methane would also provide crucial information for new interpretations having a tremendous impact on current knowledge of atmospheric chemistry, cluster and planetary science, hydrocarbon plasma, quantum chemistry models and, possibly, new exotic metastable molecular species.

[1] L.V. Trepka and H Neurt. Uberdie entstehung von negativen ionen. Zeitschrift fr Naturforschung A, 18:1295, 1963.
[2] T. E. Sharp and J. T. Dowell. Isotope effects in dissociative attachment of electrons in methane. J. Chem. Phys., 46:1530–1531, 1967.
[3] F.K. Botz and R.E. Glick. Methane temporary negative ion resonances. Chem. Phys. Lett., 33:279–283, 1975.
[4] S. R. Hunter, J. G. Carter, and L. G. Christophorou. Electron transport measurements in methane using an improved pulsed Townsend technique. J. Appl. Phys., 60:24–35, 1986.
[5] CRC Handbook of Chemistry and Physics, volume 47. CRC Press, Boca Raton, 1986-1987.
[6] M.C. Bordage. Thèse de Doctorat. PhD thesis, Université Paul Sabatier de Toulouse, France, 1995.
[7] Ch. Hollenstein, W. Schwarzenbach, A. A. Howling, C. Courteille, J.-L. Dorier, and L. Sansonnens. Anionic clusters in dusty hydrocarbon and silane plasmas. J. Vac. Sci. Tecnol. A, 14:535–539, 1996.
[8] J. de Urquijo, C.A. Arriaga , C. Cisneros and I. Alvarez. A time-resolved study of ionization, electron attachment and positive-ion drift in methane. J. Phys. D., 32:41–45, 1999.
[9] J Winter, J Berndt, S-H Hong, E Kovacevic, I Stefanovic, and O Stepanovic. Dust formation in Ar/$CH_4$ and Ar/$C_2H_2$ plasmas. Plasma Sour Sci Tech, 18:034010, 2009.
[10] E.M. Hernández, L. Hernández, C. Martínez-Flores, N. Trujillo, M. Salazar, A. Chavez, and G. Hinojosa. Electron detachment cross sections of $CH_4^-$ colliding with $O_2$ and $N_2$ below 10 keV energies. Plasma Sour Sci Tech 23:015018, 2014.
[11] M.A. Cordiner and S.B. Charnley. Negative ion chemistry in the coma of comet 1P/Halley. Meteoritics & Planetary Science, 49:21–27, 2014.
[12] G. Horvath, J.D. Skalny, N.J. Mason, M. Klas, M. Zahoran, R. Vladoiu, and M. Manole. Corona discharge experiments in admixtures of $N_2$ and $CH_4$: a laboratory simulation of Titan's atmosphere. Plasma Sour Sci Tech, 18:034016, 2009.
[13] R. C. Fortenberry. Interstellar anions: The role of Quantum Chemistry. J. Phys. Chem. A, 119:9941–9953, 2015.
[14] T.J. Millar, C. Walsh, and T. A. Field. Negative ions in Space. Chem. Rev. 117:1765-1795, 2017.
[15] B.S. Jursic. Theoretical investigation of structures and energies of the protonated methane radical cation and ethane. J. Mol. Struc.-THEOCHEM, 498:149 – 157, 2000.
[16] Although Botz and Glick [3] proposed other excited doublet states for less symmetric non-$T_d$ methane geometries as possible candidates to explain the observed anionic species, we have verified that unrestricted geometry optimizations of these excited states at the MP2 level always lead to the $^2A_1$ unstable ground state of the anion.
[17] Kirk A. Peterson, David E. Woon, and Thom H. Dunning. Benchmark calculations with correlated molecular wave functions. IV. The classical barrier height of the H + $H_2$ → $H_2$ + H reaction. J. Chem. Phys., 100:7410–7415, 1994.
[18] T.H. Dunning. Gaussian basis sets for use in correlated molecular calculations. I. The atoms boron through neon and hydrogen. J. Chem. Phys., 90:1007–1023, 1989.
[19] A. Ramírez-Solís. On the accuracy of the complete basis set extrapolation for anionic systems: A case study of the electron affinity of methane. Comput. Chem., 2:31, 2014.
[20] C.F. Bunge, M. Galán, R. Jáuregui and A.V. Bunge. Systematic search of excited states of negative ions lying above the ground state of the neutral atom. Nucl. Instr. & Methods, 202:299–305, 1982.
[21] J.W. Hiby. Massenspektrographische untersuchungen an wasserstoff- und heliumkanalstrahlen ($H_3^+$, $H_2^-$, $HeH^+$, $HeD^+$, $HeV^-$). Annalen der Physik, 426:473–487, 1939.

[22] A. Wolf, K. G. Bhushan, I. Ben-Itzhak, N. Altstein, D. Zajfman, O. Heber, and M. L. Rappaport. Lifetime measurement of He$^-$ using an electrostatic ion trap. Phys. Rev. A, 59:267–270, 1999.

[23] P. Feldman and R. Novick. Autoionizing states in alkali atoms with μs lifetimes. Phys. Rev. Lett., 11:278–281, 1963.

[24] P. Feldman and R. Novick. Autoionizing states in alkali atoms with μs lifetimes. Phys. Rev., 160:143–158, 1967.

[25] D.E. Holmgren, D.J. Walker, D.A. King, and S.E. Harris. Laser spectroscopy of Na I quartets. Phys. Rev. A, 31:677–683, 1985.

[26] C. Raynaud, L. Maron, J-P. Daudey, and F. Jolibois. Reconsidering Car-Parrinello molecular dynamics using direct propagation of molecular orbitals developed upon gaussian type atomic orbitals. Phys. Chem. Chem. Phys., 6:4226–4232, 2004.

[27] A. Ramírez-Solís, C.O. Bartulovich, Tesia V. Chciuk, J. Hernández-Cobos, H. Saint-Martin, L. Maron, W.R. Anderson, A.M. Li, and R.A. Flowers. Experimental and theoretical studies on the implications of halide-dependent aqueous solvation of Sm(II). J. Am. Chem. Soc., 140:16731–16739, 2018.